\documentclass[final,3p,times,twocolumn]{elsarticle}
\usepackage{amsmath}
\usepackage{amsfonts}
\usepackage{amssymb}
\usepackage{graphicx}
\usepackage{dcolumn}
\usepackage{color}

\begin{document}

\begin{frontmatter}

\title{Bound systems of interacting electrons. A step beyond density
functional theory.}

\author[ufterrae]{Miguel Lagos}
\ead{mlagos@gmail.com}
\address[ufterrae]{Facultad de Ingenier{\'\i}a, Universidad Finis Terrae,
Av. Pedro de Valdivia 1509, Providencia, Regi{\'o}n Metropolitana, Chile}

\begin{abstract}
The non--relativistic interacting electron gas in an external field of
positively charged massive cores is dealt with in the scheme of second
quantization. Ladder operators that change between stationary states of
contiguous energy eigenvalues are derived. The method is particularized
to the two--electron Helium atom in order to explain it avoiding too much
notation. Applying the lowering operator on the ground state must give
zero because no state with lower energy does exist. Equations for the
ground state and ground state energy are obtained this way and solved,
giving closed--form expressions for the ground state, its energy and
electronic density of Helium. The theory in its lowest order gives 0.63\%
error. The application to more complex systems and higher degrees of
approximation seems straightforward. The foundations of the density
functional theory and how to go beyond it are seen quite clearly.
\end{abstract}


\end{frontmatter}

\date{\today}

\vskip 1cm

The study of the electronic structure of molecules and solids has been
greatly facilitated by the breakthrough of the density functional theory
(DFT). By demonstrating that the ground state energy of the $N$ electrons
system in a external field can be regarded as a unique functional of the
electron density, Hohenberg and Kohn \cite{HohenbergKohn} found a bypass
to the problem of solving the many--body Schr{\"o}dinger equation that
reduces it, in principle, to the search for the density functional that
minimizes the energy. Instead of seeking for a wave function of $3N$ real
variables, the problem is substituted by determining a function of just 3
variables. Afterwards, Kohn and Sham \cite{KohnSham} applied these ideas
to a conveniently defined non--interacting system yielding the same
density as the original problem of interacting electrons
\cite{Salomonson}.

The procedure has proven to give valuable information on the stationary
structure of atoms, molecules and solids in ground state, but a complete
knowledge of the dynamical state of the system demands knowing the state
vector. This communication introduces a method that may compete with the
DFT in simplicity and accuracy, but gives also the state vector in the
second quantized formalism. The derivations are restricted here to
systems of two energy levels or bands, as are the $n=0$ and $n=1$
states of He, bonding and anti--bonding states of H$_2$, or the $\sigma$
and $\pi$ bonding orbitals of conjugate polymers. The application to
polymers and more complex systems is left for a forthcoming communication
because the study of the two--electron systems is enough to explain the
method. Just to show the power of the method the ground state vector is
used to calculate the density and excitation energy. As an interesting
byproduct of the present approach, it gives an alternate demonstration of
Hohenberg--Kohn theorem and Kohn--Sham procedure of regarding the system
as an ensemble of free fermions in effective potentials.
  
In general, the $N$--electron system whose Hamiltonian operator in
coordinate space is

\begin{equation}
H=\sum_{i=1}^N\left( -\frac{\hbar^2}{2m}\nabla_i^2+
V(\vec{r_i})\right)+\sum_{i<j}U(\vec{r_i},\vec{r_j})
\label{A1}
\end{equation}

\noindent
has a representation

\begin{equation}
H=\sum_k\epsilon_k c_k^\dagger c_k +
\sum_{klmn}U_{klmn} c_k^\dagger c_l^\dagger c_m c_n ,
\label{A2}
\end{equation}

\noindent
in the Fock space of states \cite{Roman}. The fermion operators $c_k$ are
the coefficients of the field operator

\begin{equation}
\Psi (\vec{r},t)=\sum_k c_k(t)u_k(\vec{r})
\label{A3}
\end{equation}

\noindent
expanded in the complete set of orthonormal functions ${u_k(\vec{r})}$,
defined by the solutions of the eigenvalue equation 

\begin{equation}
\left( -\frac{\hbar^2}{2m}\nabla^2+V(\vec{r})\right)\, u_k(\vec{r})
=\epsilon_k u_k(\vec{r}) .
\label{A4}
\end{equation}

\noindent
The off--diagonal coefficients are

\begin{equation}
U_{klmn}=\int\, d^3\vec{r}\, u_k^*(\vec{r})u_l^*(\vec{r'})
U(\vec{r},\vec{r'})u_m(\vec{r'})u_n(\vec{r})
\label{A5}
\end{equation}

\noindent
and the indices $k$, $l$, $m$ and $n$ incorporate the spin indices
$\uparrow$ and $\downarrow$. The occupancy of the one--electron state $k$
is $n_k=c_k^\dagger c_k$, which has the only eigenvalues 0 and 1 as the
operators $c_k$ obey Fermi--Dirac anti--commutation rules,
$\{ c_k,c_{k'}\}=0$ and $\{ c_k,c_{k'}^\dagger\} =\delta_{k{k'}}$.

Notice that the restriction $i<j$ in Eq.~(\ref{A1}) for $H$ in coordinate
space, precluding double interactions between particles, is absent in the
Fock space formulation. The condition $i<j$ follows directly from
the demonstration of the equivalence of the two representations. All
interaction terms are allowed in the second quantized formalism, and the
commutation properties of the creation and annihilation operators will
determine what of them are relevant in each situation. This is important
for the correct counting of the interaction terms.
 
Retaining in the one--electron basis set $\{ u_k\}$ only the states of at
most the two lowest energy levels, $u_{1\uparrow}$, $u_{1\downarrow}$,
$u_{2\uparrow}$ and $u_{2\downarrow}$, and assuming no spin interactions,
the Hamiltonian operating in a Fock space of states with zero total spin
becomes 

\begin{equation}
\begin{aligned}
H=\, &\epsilon_1(n_{1\uparrow}+n_{1\downarrow})+
\epsilon_2(n_{2\uparrow}+n_{2\downarrow})\\
&+V_1 n_{1\uparrow}n_{1\downarrow}+V_2 n_{2\uparrow}n_{2\downarrow}
+U(n_{1\uparrow}n_{2\downarrow}+n_{1\downarrow}n_{2\uparrow})\\
&+\overline{U}(c_{1\uparrow}^\dagger c_{1\downarrow}^\dagger
c_{2\downarrow}c_{2\uparrow}
+c_{2\uparrow}^\dagger c_{2\downarrow}^\dagger c_{1\downarrow}c_{1\uparrow}+
c_{1\uparrow}^\dagger c_{2\downarrow}^\dagger c_{1\downarrow}c_{2\uparrow}\\
&+c_{2\uparrow}^\dagger c_{1\downarrow}^\dagger c_{2\downarrow}c_{1\uparrow}),
\label{A6}
\end{aligned}
\end{equation}

\noindent
where, in Gauss units,

\begin{equation}
V_1=\int d^3\vec{r}d^3\vec{r'}\, |u_1(\vec{r})|^2|u_1(\vec{r'})|^2
\frac{e^2}{|\vec{r}-\vec{r'}|},
\label{A7}
\end{equation}

\begin{equation}
V_2=\int d^3\vec{r}d^3\vec{r'}\, |u_2(\vec{r})|^2|u_2(\vec{r'})|^2
\frac{e^2}{|\vec{r}-\vec{r'}|},
\label{A8}
\end{equation}

\noindent
$U$ and $\overline{U}$ are the correlation and exchange integrals

\begin{equation}
U=\int d^3\vec{r}\, d^3\vec{r'}\, u_1^*(\vec{r})u_2^*(\vec{r'})
\frac{e^2}{|\vec{r}-\vec{r'}|}u_2(\vec{r'})u_1(\vec{r}),
\label{A9}
\end{equation}

\noindent
and

\begin{equation}
\overline{U}=\int d^3\vec{r}\, d^3\vec{r'}\, u_1^*(\vec{r})u_2^*(\vec{r'})
\frac{e^2}{|\vec{r}-\vec{r'}|}u_1(\vec{r'})u_2(\vec{r}).
\label{A10}
\end{equation}

The method will be introduced by calculating a well known system of
two interacting electrons, the helium atom. In the scheme settled before
the one--electron basis set of functions is constrained to just the $s$
and $p$ one--electron hydrogenic states

\begin{equation}
u_{1s}(\vec{r})=\frac{1}{\sqrt{\pi a^3}}
\exp{\left( -\frac{r}{a}\right)}\chi_s
\quad \epsilon_1=-\frac{2me^4}{\hbar^2}
\quad a=\frac{\hbar^2}{2me^2}
\label{A11}
\end{equation}

\noindent
and

\begin{equation}
u_{2s}(\vec{r})=\frac{1}{2\sqrt{2\pi a^3}}
\left( 1-\frac{r}{2a}\right)\exp{\left( -\frac{r}{2a}\right)}\chi_s
\quad \epsilon_2=-\frac{me^4}{2\hbar^2} ,
\label{A12}
\end{equation}

\noindent
where $\chi_s$, $s=\uparrow ,\,\downarrow$, denotes spin eigenstates and 
$m$, $e$ are the electron mass and fundamental charge.

The Hamiltonian of the helium atom in Born--Oppenheimer approximation and
standard notation

\begin{equation}
H=-\frac{\hbar^2}{2m}\nabla_1^2-\frac{\hbar^2}{2m}\nabla_2^2
-\frac{2e^2}{r_1}-\frac{2e^2}{r_2}+\frac{e^2}{|\vec{r}_2-\vec{r}_1|}
\label{A13}
\end{equation}

\noindent
in second quantization formalism reads

\begin{equation}
\begin{aligned}
H=\,&\epsilon_1(n_{1\uparrow}+n_{1\downarrow})+\epsilon_2(n_{2\uparrow}
+n_{2\downarrow})+V_1 n_{1\uparrow}n_{1\downarrow} \\
&+V_2 n_{2\uparrow}n_{2\downarrow}
+U(n_{1\uparrow}n_{2\downarrow}+n_{1\downarrow}n_{2\uparrow})\\
&+\overline{U}(\tilde{\psi}+\tilde{\psi}^\dagger
-\tilde{\phi}-\tilde{\phi}^\dagger ),
\end{aligned}
\label{A14}
\end{equation}

\noindent
where the operators $\tilde{\psi}$ and $\tilde{\phi}$ are defined by

\begin{equation}
\tilde{\phi} =
c_{1\uparrow}^\dagger c_{2\downarrow}^\dagger c_{2\uparrow}
c_{1\downarrow} \qquad
\tilde{\psi} =
c_{1\uparrow}^\dagger c_{1\downarrow}^\dagger c_{2\downarrow}
c_{2\uparrow}.
\label{A15}
\end{equation}

Making repeated use of the elementary identities

\begin{equation}
\begin{aligned}
&[AB,C]=A[B.C]+[A,C]B,\\
&[A,BC]=[A,B]C+B[A,C], \\
&[AB,C] =A\{ B.C\} -\{ A,C\} B,\\
&[A,BC] =\{ A,B\} C-B\{ A,C\} ,
\end{aligned}
\label{A16}
\end{equation}

\noindent
it can be shown that the operators (\ref{A15}) have the commutation
properties

\begin{equation}
\begin{aligned}
&[\tilde{\phi},\tilde{\phi}^\dagger]=
-n_{1\uparrow}n_{2\downarrow}(1-n_{1\downarrow}-n_{2\uparrow})
+n_{1\downarrow}n_{2\uparrow}(1-n_{1\uparrow}-n_{2\downarrow})\\
&[\tilde{\psi},\tilde{\psi}^\dagger]=
n_{1\uparrow}n_{1\downarrow}(1-n_{2\uparrow}-n_{2\downarrow})
-n_{2\uparrow}n_{2\downarrow}(1-n_{1\uparrow}-n_{1\downarrow}).
\label{A17}
\end{aligned}
\end{equation}

\noindent
One can also demonstrate that

\begin{equation}
\begin{aligned}
&[n_{1\uparrow},\tilde{\psi}^\dagger]
=[n_{1\downarrow},\tilde{\psi}^\dagger]
=-\tilde{\psi}^\dagger ,\\
&[n_{2\uparrow},\tilde{\psi}^\dagger]
=[n_{2\downarrow},\tilde{\psi}^\dagger]
=\tilde{\psi}^\dagger ,
\end{aligned}
\label{A18}
\end{equation}

\noindent
and

\begin{equation}
\begin{aligned}
&[n_{1\uparrow}n_{1\downarrow},\tilde{\psi}^\dagger]=
-\tilde{\psi}^\dagger (n_{1\uparrow}+n_{1\downarrow}-1) \\
&\phantom{ababababab}
=-(n_{1\uparrow}+n_{1\downarrow}+1)\tilde{\psi}^\dagger \\
&[n_{2\uparrow}n_{2\downarrow},\tilde{\psi}^\dagger]=
\tilde{\psi}^\dagger (n_{2\uparrow}+n_{2\downarrow}+1) \\
&\phantom{ababababab}
=(n_{2\uparrow}+n_{2\downarrow}-1)\tilde{\psi}^\dagger .
\end{aligned}
\label{A19}
\end{equation}

\noindent
Moreover,

\begin{equation}
\begin{aligned}
&[n_{1\uparrow}n_{2\uparrow},\tilde{\psi}^\dagger]=
\tilde{\psi}^\dagger (n_{1\uparrow}-n_{2\uparrow}-1)=
(n_{1\uparrow}-n_{2\uparrow}+1)\tilde{\psi}^\dagger , \\
&[n_{1\downarrow}n_{2\downarrow},\tilde{\psi}^\dagger]=
\tilde{\psi}^\dagger (n_{1\downarrow}-n_{2\downarrow}-1)=
(n_{1\downarrow}-n_{2\downarrow}+1)\tilde{\psi}^\dagger , \\
&[n_{1\uparrow}n_{2\downarrow},\tilde{\psi}^\dagger]=
\tilde{\psi}^\dagger (n_{1\uparrow}-n_{2\downarrow}-1)=
(n_{1\uparrow}-n_{2\downarrow}+1)\tilde{\psi}^\dagger , \\
&[n_{1\downarrow}n_{2\uparrow},\tilde{\psi}^\dagger]=
\tilde{\psi}^\dagger (n_{1\downarrow}-n_{2\uparrow}-1)=
(n_{1\downarrow}-n_{2\uparrow}+1)\tilde{\psi}^\dagger , \\
\end{aligned}
\label{A20}
\end{equation}

\noindent
and

\begin{equation}
[\tilde{\phi},\tilde{\psi}^\dagger]=
[\tilde{\phi}^\dagger ,\tilde{\psi}^\dagger]=0 ,
\label{A21}
\end{equation}

\noindent
together with a series of similar equations for the $\phi$--operators,
e.~g. $[n_{1\uparrow},\tilde{\phi}]=-[n_{2\uparrow},\tilde{\phi}]=
-[n_{1\downarrow},\tilde{\phi}]=[n_{2\downarrow},\tilde{\phi}]=
\tilde{\phi}$.

From these commutation properties it readily follows that 

\begin{equation}
\begin{aligned}
&[H,\tilde{\psi}^\dagger ]=2\big[ -\epsilon_1 +\epsilon_2
-V_1(n_{1\uparrow}+n_{1\downarrow}+1)\\
&+V_2(n_{2\uparrow}+n_{2\downarrow}-1)\\
&+U(n_{1\uparrow}+n_{1\downarrow}
-n_{2\uparrow}-n_{2\downarrow}+2)\big]\tilde{\psi}^\dagger\\
&+2\overline{U}\big[ n_{1\uparrow}n_{1\downarrow}
(1-n_{2\uparrow}-n_{2\downarrow})
-n_{2\uparrow}n_{2\downarrow}(1-n_{1\uparrow}-n_{1\downarrow})\big],
\end{aligned}
\label{A22}
\end{equation}

\begin{equation}
\begin{aligned}
&[H,\tilde{\psi}]=2\big[ \epsilon_1 -\epsilon_2
+V_1(n_{1\uparrow}+n_{1\downarrow}-1)\\
&-V_2(n_{2\uparrow}+n_{2\downarrow}+1)\\
&-U(n_{1\uparrow}+n_{1\downarrow}
-n_{2\uparrow}-n_{2\downarrow}-2)\big]\tilde{\psi}\\
&-2\overline{U}\big[ n_{1\uparrow}n_{1\downarrow}
(1-n_{2\uparrow}-n_{2\downarrow})
-n_{2\uparrow}n_{2\downarrow}(1-n_{1\uparrow}-n_{1\downarrow})\big],
\end{aligned}
\label{A23}
\end{equation}

\noindent
where care was taken in Eqs.~(\ref{A22}) and (\ref{A23}) of placing the
$\psi$--operators at the right side.

Notice that all the commutators in the preceding sub--section have the
remarkable property of being expressed in terms of just the occupation
numbers $n_{1s}$ and $n_{2s}$ of the truncated one--electron basis.
Although the occupation numbers are in general not conserved magnitudes,
their mean values will be taken here as approximately good quantum
numbers. More precisely, it will be assumed that when the number
operators $n_{k}$ are applied directly to a stationary state of the
system they can be replaced by their classical expectation value. These
expectation values are such that they make extremal the mean energy of
the system. The scheme proposed here for the He atom is particularly
simple because we have only four basis states and the occupation numbers
are not expected to depend on the spin. Hence one can define

\begin{equation}
\langle n_{1\uparrow}\rangle =\langle n_{1\downarrow}\rangle = 1-\eta,
\qquad
\langle n_{2\uparrow}\rangle =\langle n_{2\downarrow}\rangle = \eta,
\label{A24}
\end{equation}

\noindent
where $0\le\eta\le 1$ is a unique variational parameter. The commutators
(\ref{A22}) and (\ref{A23}) then become

\begin{equation}
\begin{aligned}
&[H,\tilde{\psi}^\dagger ]=2\big[ -\epsilon_1+\epsilon_2-V_1(3-2\eta)-
V_2(1-2\eta)\\
&\phantom{ababab}
+4U(1-\eta)\,\big]\tilde{\psi}^\dagger
+2\overline{U}(1-2\eta)(1-2\eta+2\eta^2)
\end{aligned}
\label{A25}
\end{equation}

\noindent
and

\begin{equation}
\begin{aligned}
&[H,\tilde{\psi}]=2\big[ \epsilon_1 -\epsilon_2
+V_1(1-2\eta)-V_2(1+2\eta)\\
&\phantom{abababab}
+4U\eta\big]\tilde{\psi} -2\overline{U}(1-2\eta)(1-2\eta+2\eta^2).
\end{aligned}
\label{A26}
\end{equation}

\noindent
Defining now the two operators

\begin{equation}
\psi_{+}^\dagger =\tilde\psi^\dagger +\Lambda_{+}
\quad\text{and}\quad 
\psi_{-} =\tilde\psi +\Lambda_{-}
\label{A27}
\end{equation}

\noindent
where

\begin{equation}
\Lambda_{+} =
\frac{-(1-2\eta)(1-2\eta+2\eta^2)\overline{U}}{-\epsilon_1+
\epsilon_2-V_1(3-2\eta)-V_2(1-2\eta)+4U(1-\eta)}\, ,
\label{A28}
\end{equation} 

\begin{equation}
\Lambda_{-}
=\frac{(1-2\eta)(1-2\eta+2\eta^2)\overline{U}}{\epsilon_1-
\epsilon_2+V_1(1-2\eta)-V_2(1+2\eta)+4U\eta}\, ,
\label{A29}
\end{equation}

\noindent
the commutators (\ref{A25}) and (\ref{A26}) become

\begin{equation}
\begin{aligned}
&[H,\psi_{+}^\dagger ]=2[-\epsilon_1+\epsilon_2-V_1(3-2\eta)-
V_2(1-2\eta)\\
&\phantom{abababab}+4U(1-\eta)]\psi_{+}^\dagger
\end{aligned}
\label{A30}
\end{equation}

\noindent
and

\begin{equation}
\begin{aligned}
&[H,\psi_{-}]=
2\big[\epsilon_1 -\epsilon_2 +V_1(1-2\eta)-V_2(1+2\eta)\\
&\phantom{abababab}+4U\eta
\big]\psi_{-}.
\end{aligned}
\label{A31}
\end{equation}

\noindent
This way the operators $\psi_{+}^\dagger$ and $\psi_{-}$ are raising
and lowering operators. They are not just conjugate operators because
the energy levels of the system are not equally spaced, as is for the
harmonic oscillator. If $H|n\rangle =E_n|n\rangle$ then vectors
$\psi_{+}^\dagger |n\rangle$ and $\psi_{-} |n\rangle$ are also
eigenvectors of $H$ with energy eigenvalues 

\begin{equation}
E_{n+1}=E_n +2\big[ -\epsilon_1+\epsilon_2-V_1(3-2\eta)-V_2(1-2\eta)
+4U(1-\eta)\big]
\label{A32}
\end{equation}

\noindent
and

\begin{equation}
E_{n-1}=E_n+2\big[\epsilon_1-\epsilon_2 +V_1(1-2\eta)
-V_2(1+2\eta)-4U\eta\big].
\label{A33}
\end{equation}

If $[\psi ,\psi^\dagger ]=K$, for real $K$ and being $f$ an analytic
function whose derivative is $f'$, then it holds the identity

\begin{equation}
\left[\psi ,f\left(\frac{\psi^\dagger -\psi}{K}\right)\right]
=f'\left(\frac{\psi^\dagger -\psi}{K}\right),
\qquad K \in \mathbb{R}.
\label{A34}
\end{equation}

\noindent
To demonstrate this by complete induction consider first the simpler
case of two operators $A$ and $B$ such that $[A,B]=1$ and show that
$[A,B^n]=nB^{(n-1)}$. The case $n=1$ is trivially fulfilled because
the relation to be demonstrated reduces to the hypothesis. Hence it
only rests to show that if the thesis holds for an arbitrary value of
$n$ then it also holds for $n+1$. With this aim recall Eqs.~(\ref{A16})
to show that $[A,B^{(n+1)}]=B[A,B^n]+[A,B]B^n$. This reduces to
$[A,B^{(n+1)}]=(n+1)B^n$ when replacing the induction hypotheses
$[A,B]=1$ and $[A,B^n]=nB^{(n-1)}$, which completes the proof. The
property can be extended directly to $[A,f(B)]=f'(B)$. Eq.~(\ref{A34})
follows from substituting $A=\psi$ and $B=(\psi^\dagger -\psi)/K$.

From Eqs.~(\ref{A17}), (\ref{A24}), and (\ref{A27}) one has that

\begin{equation}
[\psi_{+} ,\psi_{+}^\dagger]=[\psi_{-} ,\psi_{-}^\dagger ]=
(1-2\eta)(1-2\eta +2\eta^2)\equiv K.
\label{A35}
\end{equation}

\noindent
Hence, applying Eq.~(\ref{A33}) we have that

\begin{equation}
\psi_{\pm}\exp\bigg[ -\frac{\Lambda_{\pm}}{K}(\psi_{\pm}^\dagger -\psi_{\pm} )\bigg] =
\exp\bigg[ -\frac{\Lambda_{\pm}}{K}(\psi_{\pm}^\dagger -\psi_{\pm} )\bigg]\,
(\psi_{\pm} -\Lambda_{\pm}) .
\label{A36}
\end{equation}

The ground state $|g\rangle $ follows the equation

\begin{equation}
\psi_{-}|g\rangle =0
\label{A37}
\end{equation}

\noindent
because no state exists with energy below the energy eigenvalue $E_g$ of
$|g\rangle $. To find the solution of Eq.~(\ref{A37}) recall
Eqs.~(\ref{A15}) and (\ref{A27}) to show that $\psi_{-}-
\Lambda_{-}=c_{1\uparrow}^\dagger c_{1\downarrow}^\dagger
c_{2\downarrow}c_{2\uparrow}$. Combining this with  Eq.~(\ref{A27})
for $\psi_{-}$ one has that

\begin{equation}
\begin{aligned}
&\psi_{-}\exp \bigg[ -\frac{\Lambda_{-}}{K}(\psi_{-}^\dagger -\psi_{-} )\bigg]
c_{1\uparrow}^\dagger c_{1\downarrow}^\dagger |0\rangle \\
&=\exp\bigg[ -\frac{\Lambda_{-}}{K}(\psi_{-}^\dagger -\psi_{-} )\bigg]\,
c_{1\uparrow}^\dagger c_{1\downarrow}^\dagger c_{2\downarrow} c_{2\uparrow}
c_{1\uparrow}^\dagger c_{1\downarrow}^\dagger |0\rangle =0
\label{A38}
\end{aligned}
\end{equation}

\noindent
by the nilpotent character of the $c$--operators. Thus,

\begin{equation}
|g\rangle =\exp\left[ -\frac{\Lambda_{-}}{K}
(\psi_{-}^\dagger -\psi_{-})\right]
c_{1\uparrow}^\dagger c_{1\downarrow}^\dagger |0\rangle .
\label{A39}
\end{equation}

\noindent
Writting Eq.~(\ref{A39}) in terms of the electron operators the ground
state takes the form

\begin{equation}
\begin{aligned}
|g\rangle =&\exp\bigg(
\frac{(c_{1\uparrow}^\dagger c_{1\downarrow}^\dagger
c_{2\downarrow} c_{2\uparrow}-
c_{2\uparrow}^\dagger c_{2\downarrow}^\dagger
c_{1\downarrow} c_{1\uparrow})\overline{U}}
{\epsilon_1 -\epsilon_2 +V_1(1-2\eta)-V_2(1+2\eta)+4U\eta}\bigg)\\
&\times c_{1\uparrow}^\dagger c_{1\downarrow}^\dagger |0\rangle .
\end{aligned}
\label{A40}
\end{equation}

Up to this point only the $\psi$--operators were taken into account. An
entirely similar procedure can be applied to the $\phi$--operators, which
yields the equation $[H,\tilde\phi]=(V_1-V_2+2U(1-\eta))\tilde\phi$. The
$\phi$--operators prove not to contribute significantly to the ground
state.

The expectation value of the Hamiltonian (\ref{A14}) for the ground state
turns out to be

\begin{equation}
\begin{aligned}
&E_g =\langle g|H|g\rangle =\, 2(1-\eta)\epsilon_1+2\eta\epsilon_2 \\
&+2V_1 (1-\eta)^2+2V_2\eta^2+4U\eta (1-\eta)
-\overline{U}(\Lambda_{+}+\Lambda_{-})
\label{A41}
\end{aligned}
\end{equation}

\noindent
But for the last one, the terms in the right hand side of this equation
are easily derived from Eq.~(\ref{A14}) for $H$, just replacing
Eqs.~(\ref{A24}) and (\ref{A39}). The last term $\langle g|(\tilde\psi +
\tilde\psi^\dagger -\tilde\phi^\dagger -\tilde\phi )|g\rangle$
demands a little bit more attention. The operators $\tilde\phi$ and
$\tilde\phi^\dagger$ commute with $\tilde\psi$ and $\tilde\psi^\dagger$.
Hence they pass through the exponential in Eq.~(\ref{A39}) and
annihilate particles in the vacuum. Their contributions then vanish.
This way the only contributions to the bracket come from the operators
$\tilde\psi$ and $\tilde\psi^\dagger$, which lend factors $-\Lambda_{-}$
and $-\Lambda_{+}$ each, as indicated in Eqs.~(\ref{A27}), (\ref{A28})
and (\ref{A29}). The value of $\eta$ is the one which minimizes
$\langle g|H|g\rangle$.

To have a test of the accuracy of the procedure, it is applied to the
Helium atom. The basis functions are the hydrogenic $1s$ and $2s$
wavefunctions (\ref{A11}) and (\ref{A12}), where $a$ is twice Bohr's
radius. Just the sub--space of zero total angular momentum is considered.
Solving the integrals (\ref{A7})--(\ref{A10}) one has that

\begin{equation}
\begin{aligned}
&V_1=-2\frac{e^2}{a},\quad V_2=-\frac{e^2}{2a},\\
&U=\frac{17}{162}\frac{e^2}{a},
\quad\overline{U}=\frac{8}{729}\frac{e^2}{a}.
\end{aligned}
\label{A42}
\end{equation}

\noindent
The coefficient of $\overline{U}$, and $\overline{U}$ itself, turns out
to be very small. As it appears as a multiplicative factor in expressions
(\ref{A28}) and (\ref{A29}), the contribution of the term
$\Lambda_{+}+\Lambda_{-}$ to $E_g$ is of second order in $\overline{U}$
and can be neglected for the purpose of calculating $\eta$. This is a
very interesting point because all the other terms of $E_g$ are
proportional to coefficients that can be obtained from just the
electronic density. Hence the basic hypothesis of Hohenberg and Kohn is
clearly shown in the most basic case of the Helium atom. The ground state
energy is then

\begin{equation}
\begin{aligned}
E_g=\,&2\epsilon_1+2V_1+(-2\epsilon_1+2\epsilon_2-4V_1+4U)\eta \\
&+2(V_1+V_2-2U)\eta^2.
\end{aligned}
\label{A43}
\end{equation}

Replacing $\epsilon_1 =-e^2/a$ and $\epsilon_2 =-e^2/(4a)$, as given by
Eqs.~(\ref{A11}) and (\ref{A12}), it is readily shown that the resulting
expression reaches a minimum 

\begin{equation}
\begin{aligned}
E_g &= -1.4610\,\frac{e^2}{a}\,[\text{Gauss}] \\
&=-2.9220\,[\text{a.u.}]=-79.51\,[\text{eV}]
\end{aligned}
\label{A44}
\end{equation}

\noindent
for the occupation number $\eta =0.91515$. which is the value that
minimizes Eq.~(\ref{A43}). This result should be compared with the
calculation of Korobov \cite{Korobov}, which incorporates as many as 5200
basis functions and gives the figure $E_g =-2.90372... \,[\text{a.u.}]$.
Hence our result (\ref{A44}) differs by 0.63\% from the exact, employing
just two basis functions. The experimental value is
$-2.9034 [\text{a.u.}]$ and the Hartree--Fock limit is
$-2.8617\,[\text{a.u.}]$ \cite{BasedenTye}.

The method was restricted here to Helium, the most simple and well known
system of interacting electrons \cite{Korobov,Li}, discarding any
complication aimed to attain accuracy, just to introduce it in the most
clear way. To reach better accuracy one has simply to introduce more
basis functions and take the rather tedious job of calculating the
integral coefficients $V$, $U$ and $\overline{U}$.

Just to give an example of how to use these results the ground state
density $\rho (\vec{r})=\langle g| \Psi^\dagger(\vec{r})\Psi (\vec{r})
|g\rangle$ is evaluated next. Making use of

\begin{equation}
(\psi_{-}^\dagger -\psi_{-})^m
c_{1\uparrow}^\dagger c_{1\downarrow}^\dagger |0\rangle =
\begin{cases}
&(-1)^{m/2}c_{1\uparrow}^\dagger c_{1\downarrow}^\dagger |0\rangle ,
\qquad m\text{ even} \\
&(-1)^{(m+1)/2}c_{2\uparrow}^\dagger c_{2\downarrow}^\dagger |0\rangle ,
\quad m\text{ odd} ,
\end{cases}
\label{A45}
\end{equation}

\noindent
and Eq.~(\ref{A39}) one can easily show that

\begin{equation}
\rho (\vec{r})=
2\cos^2\bigg(\frac{\Lambda_{-}}{K}\bigg) |u_1(\vec{r})|^2+
2\sin^2\bigg(\frac{\Lambda_{-}}{K}\bigg) |u_2(\vec{r})|^2 ,
\label{A46}
\end{equation}

\noindent
where $u_1(\vec{r})$ and $u_2(\vec{r})$ are given by Eqs.~(\ref{A11}) and
(\ref{A12}).

\end{document}